\documentclass[aps,prd,twocolumn,amssymb,amsmath,showpacs,a4paper,superscriptaddress,nofootinbib]{revtex4-1}

\usepackage{graphicx}
\usepackage{amsfonts}
\usepackage{amsmath}
\usepackage{hyperref}
\usepackage{units}
\usepackage[latin1]{inputenc}
\usepackage{color}
\usepackage{dcolumn}
\usepackage{bm}
\usepackage{float}
\usepackage{cleveref}
\usepackage[normalem]{ulem}
\usepackage{appendix}

\graphicspath{ {images/} }
\usepackage{mathrsfs}
\usepackage{amssymb}
\usepackage{dsfont}
\usepackage{enumitem}
\usepackage{gensymb}
\usepackage{bm}

\crefname{section}{§}{§§}
\Crefname{section}{§}{§§}
\usepackage{mathtools}
\usepackage{chngcntr}
\counterwithout{equation}{section}

\newcommand{\cmmnt}[1]{\ignorespaces}

\newcommand{\be} {\begin{equation}}
\newcommand{\ee} {\end{equation}}
\newcommand{\bsub}{\begin{subequations}}
\newcommand{\esub}{\end{subequations}}
\newcommand{\bea}{\begin{eqnarray}}
\newcommand{\eea}{\end{eqnarray}}

\usepackage[normalem]{ulem}

\def\be{\begin{equation}}
\def\ee{\end{equation}}
\def\ba{\begin{align}}
\def\ea{\end{align}}

\begin{document}


\title{Backreaction in an analogue black hole experiment}

\author{Sam Patrick}
 \email{sampatrick31@googlemail.com}
 \affiliation{%
 School of Mathematical Sciences, University of Nottingham\\
 Nottingham, NG7 2FD, United Kingdom
}%
\author{Harry Goodhew}%
 \email{hfg23@cam.ac.uk}
 \affiliation{%
 Institute of Astronomy, University of Cambridge\\
 Cambridge, CB3 0HA, United Kingdom
}%
\author{Cisco Gooding}%
 \email{cisco.gooding@nottingham.ac.uk}
\affiliation{%
 School of Mathematical Sciences, University of Nottingham\\
 Nottingham, NG7 2FD, United Kingdom
}%
\author{Silke Weinfurtner}%
 \email{silkiest@gmail.com}
\affiliation{%
 School of Mathematical Sciences, University of Nottingham\\
 Nottingham, NG7 2FD, United Kingdom
}%
\affiliation{
Centre for the Mathematics and Theoretical Physics of Quantum Non-Equilibrium Systems, \\
University of Nottingham, Nottingham, NG7 2FD, United Kingdom
}

\date{\today}

\begin{abstract}
    In general relativity, the interaction between a black hole and the fields around it (a process known as backreaction) proceeds via the evolution of the black holes mass and angular momentum. Analogue models of gravity, particularly fluid mechanical analogues, have been very successful in mimicking the propagation of fields, and the effects they experience, around black holes. However, hydrodynamic black holes are externally driven systems whose effective mass and angular momentum are set by experimental parameters, and as such no significant internal backreaction processes are expected to take place. We show, using a rotating draining vortex flow, that a fluid system of finite size exhibits a memory that keeps track of scattering processes in the system. This memory is encoded in the total mass of the system and hence, the backreaction arises as a significant global change in the background parameters, as opposed to a small local correction. More importantly, this backreaction is encapsulated by a dynamical metric, raising the possibility of studying wave-background interaction around evolving black hole spacetimes.  
\end{abstract}

\maketitle

\textbf{\emph{Introduction.}} Analogue gravity, pioneered by Unruh in 1981 \cite{unruh1981experimental}, is a research programme which studies gravitational phenomena from general relativity (GR) using a wide variety of non-gravitational systems (see \cite{barcelo2011analogue} for a review). 
Unruh originally considered the propagation of sound waves through a fluid, and argued that if the fluid becomes supersonic in some region, the system exhibits a dumb hole horizon - the analogue of a black hole horizon. 
More generally, he showed that the wave equation describing the propagation of linear fluctuations $\phi$ through an ideal fluid is equivalent to the Klein-Gordon (KG) equation,
\begin{equation} \label{KleinGordon}
\frac{1}{\sqrt{-g}}\partial_\mu(\sqrt{-g}g^{\mu\nu}\partial_\nu\phi) = 0,
\end{equation} 
describing the propagation of a massless scalar field on an effective curved spacetime. 
The effective metric $g_{\mu\nu}$ describing such a spacetime is completely determined by the properties of the fluid flow under consideration, raising the intriguing possibility of studying general relativistic wave phenomena in the laboratory.

Many different analogues have since been investigated in a variety of condensed matter systems \cite{rousseaux2008observation,steinhauer2014laser,steinhauer2016observation,belgiorno2010hawking,vocke2018rotating,fifer2018mimicking},
including surface waves on top of a shallow fluid \cite{schutzhold2002gravity}. 
Although the analogy was originally conceived to investigate the trans-Planckian problem associated with Hawking radiation \cite{unruh2005universality}, analogue gravity has enjoyed a number of other successes: notably surface wave experiments have been used to measure Hawking radiation \cite{euve2016observation,weinfurtner2011measurement,weinfurtner2013classical}, superradiance \cite{torres2017rotational}, and quasi-normal ringing \cite{torres2018application}.

One particularly simple model of a rotating black hole is provided by surface waves propagating on a rotating, draining fluid flow - the so-called draining bathtub vortex (DBT). 
Much work has been devoted to understanding features of this model (e.g. \cite{basak2003superresonance,basak2003reflection,cardoso2004qnm,berti2004qnm,anacleto2011superresonance,dolan2012resonances,dolan2013scattering,richartz2015rotating,dempsey2016waves,churilov2018scattering,banerjee2018bacteria}), most of which relies on the assumption of an inviscid, incompressible, irrotational fluid in shallow water.
Modifications resulting from the violation of these last two assumptions have been considered \cite{patrick2018black,torres2017waves}, 
and black hole effects (superradiance \cite{torres2017rotational} and quasinormal ringing \cite{torres2018application}) have been shown to persist under such conditions.

Another key assumption, which all analogue systems rely upon, is that the waves propagate on a fixed background.
This assumption is necessarily violated in both hydrodynamical as well as graviational systems, since the fluctuations drive the evolution of the background through non-linear terms in the equations of motion.
This process in known as the backreaction.
The usual justification for neglecting backreaction is that the non-linear terms appear at quadratic order in perturbation theory and have little influence on the fluctuations, which are studied at linear order.
However, since these terms can grow in time, they will eventually become important in determining the dynamics of the background.
In atmospheric physics, this has widely been studied under the name wave-mean interaction theory \cite{buhler2014waves} to predict large scale changes in atmospheric currents resulting from small perturbations.
In studies of surface fluctuations, it has long been recognised that waves produce a second order mass flux in the direction of wave propagation \cite{philips1977dynamics}, and that this mass flux induces a drift velocity called the Stokes drift \cite{kenyon1969stokes,hasselmann1963conservation}.
Despite recognition from the fluid dynamics community,
these effects have yet to be incorporated into the analogue gravity formalism.

In this letter, we study the backreaction in an analogue black hole simulator by scattering surface waves with a DBT vortex.
Whilst the backreaction is to be expected in any non-linear system, it normally proceeds via a small local change in the background parameters.
By contrast, we show that our system exhibits a significant global change in one of the background parameters: the total fluid mass.
We demonstrate theoretically that this can be explained by way of the wave induced mass flux across the open boundary of a finite sized system.
Due to the analogy with black hole physics, the evolving background induces changes in the effective metric, as energy and angular momentum are exchanged between the waves and the analogue black hole.
Therefore, one could in principle use a system of this type to investigate backreaction due to superradiance and Hawking radiation.

\textbf{\emph{Theory.}} The system we are considering is a stationary draining fluid flow, where water enters a tank via an inlet and exits at a drain in a continuous cycle. Such a setup has previously been employed as a simulator for black hole superradiance~\cite{torres2017rotational} and ringdown~\cite{torres2018application} processes. For simplicity, we assume cylindrical symmetry about the drain and adopt polar coordinates $(r,\theta,z)$ centred on the drain.
Water occupies a region $z\in[0,H]$ in the vertical direction. The area of the tank $\mathcal{A}$ (in the direction perpendicular to the vertical) is assumed constant and is bounded by the surface $\gamma$.
For an incompressible fluid with density $\rho_f=\mathrm{const}$, the system is fully characterised by the water height $H$ and velocity field $\mathbf{V}$.
Let $\gamma$ be comprised of the surfaces $r=r_1$, encircling (and nearby) the drain, and $r=r_2$, which forms the outer wall of the tank. We assume the inlet condition is specified on a small section of $r=r_2$, but any $\theta$ dependence introduced is confined to a small layer at the edge of the tank which we neglect.

Now consider fluctuations (i.e. surface waves) described by $(h,\mathbf{v})$, which are switched on at $t=0$ and are created inside $\mathcal{A}$.
It is well-known \cite{philips1977dynamics} that linear surface waves propagating on a background flow produce a mass flux $\mathbf{j}$ in the direction of wave propagation (which in our case is in the plane $\perp$ to the vertical).
If the background $(H,\mathbf{V})$ is stationary for $t<0$, then shortly after the onset of waves, the amount of mass $M$ contained within the system will be altered if there is a net mass flux $\mathbf{j}_\perp=\rho_f\mathbf{v}_\perp$ over $\gamma$, according to
\begin{equation} \label{Hdot0}
\dot{M} = -\int_\gamma \rho_f h\mathbf{v}_\perp\cdot d\mathbf{l},
\end{equation}
where the overdot denotes the time derivative.
The assumptions involved in the derivation of this formula are detailed in Appendix A.
Furthermore, since $M=\rho_f\int_\mathcal{A}HdA$, a change in total mass results in a change in the water height.
If we assume that $H$ is approximately level over $\mathcal{A}$ (i.e. spatially uniform), then the water height adjusts according to,
\begin{equation} \label{Hdot4}
\dot{H}_0 \simeq -\frac{1}{\mathcal{A}}\int_\gamma h\mathbf{v}_\perp\cdot d\mathbf{l},
\end{equation}
where subscript $0$ indicates that strictly a quantity is evaluated at $t=0$. Expanding $H(t)$ around $t=0$ gives,
\begin{equation} \label{Hlin}
H(t) = H_0 + \dot{H}_0t + \mathcal{O}(t^2).
\end{equation}
The $\mathcal{O}(t^2)$ corrections become significant once changes to the background become large, which means that terms quadratic in the background variables also contribute to Eq.~\eqref{Hdot4}.
Once this happens, the $t$ dependence of $H$ and $\mathbf{V}_\perp$ becomes interlinked and to solve the coupled system, one must use a second equation (the equation of momentum conservation) in addition to the equation of mass conservation which is used in the derivation of Eq.~\eqref{Hdot0}.
Although we do not attempt this here, we expect this to produce exponential behaviour of $H(t)$ at late times, since $\dot{H}$ will depend on the value of $H$.

Near $t=0$, fluctuations of frequency $\omega$ perceive a quasi-stationary background and may be written,
\begin{equation} \label{ansatz}
f(t,\theta,r,z) = \sum_m f_m(r,z) e^{im\theta-i\omega t},
\end{equation}
where $f$ is a placeholder for $(h,\mathbf{v})$ and we have also used our assumption of cylindrical symmetry to make a decomposition into modes of azimuthal number $m\in(-\infty,\infty)$.
Using this ansatz, Eq.~\eqref{Hdot4} becomes
\begin{equation} \label{Hdot5}
\dot{H}_0 = -\frac{2\pi}{\mathcal{A}}\sum_m \tfrac{1}{2}\mathrm{Re}[h_m^*\mathbf{v}_m]\cdot\mathbf{r}\big|^{r_2}_{r_1},
\end{equation}
where $*$ denotes the complex conjugate and we have dropped the subscript $\perp$ on the velocity perturbation for conciseness.
At $r=r_2$, $\mathbf{v}_m\cdot\mathrm{\mathbf{e}}_r$ vanishes everywhere except at the inlet, where we assume a rapid influx of water (exceeding the propagation speed of the fluctuations) required to drive the high flow velocities. Therefore, since the fluctuations are generated inside $\mathcal{A}$, the total mass flux receives no contribution at $r=r_2$ and is determined solely by the form of the fluctuations at $r=r_1$.
In Appendix B, we show how Eq.~\eqref{Hdot5} can be evaluated in the shallow water regime for a DBT vortex. 
This system exhibits an effective horizon at $r=r_h$ which we take as our inner boundary.

\begin{figure*} [t!] 
\centering
\includegraphics[width=\linewidth]{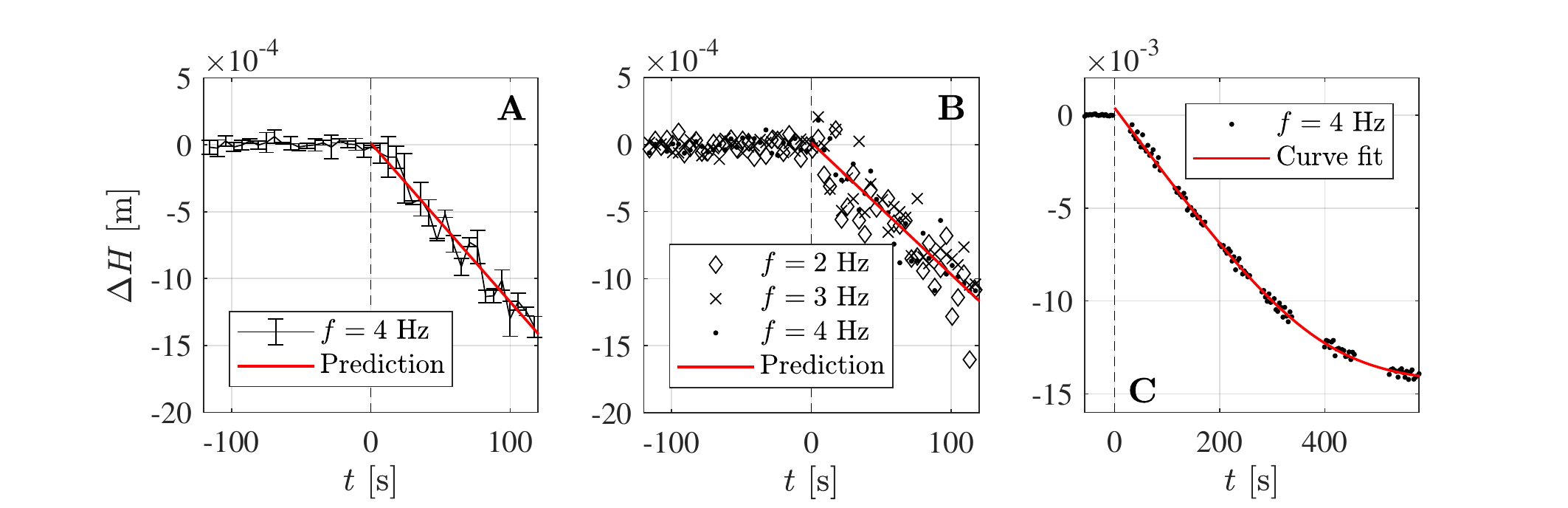}
\caption{The height change $\Delta H$ in a DBT resulting from wave incidence.
The dashed vertical line indicates when the first wavefront passed the vortex. 
    In experiment 1 (panel A), we find a consistent decrease in the water level when sending waves of frequency $f=4~\mathrm{Hz}$. We have plotted the average over three different repeats of the same experiment, and the error bars represent the standard deviation.
    In experiment 2 (panel B), we find similar behaviour when sending waves of varying frequencies.
    The model is fitted to the average of the three data sets.
    In both experiments 1 and 2 we use the linear model in Eq.~\eqref{Hlin}.
    In experiment 3 (panel C) we record the height change over a longer period of wave stimulation, finding good agreement with the heuristic fit in Eq.~\eqref{sech2} which has exponential behaviour at late times.
    Parameters obtained from the fits can be found in Table~I
    } \label{fig:results}
\end{figure*}

\textbf{\emph{Experiments.}} 
To test the prediction of Eqs.~\eqref{Hlin} and \eqref{Hdot5}, as well as to observe the time-scales over which they are valid, we scatter monochromatic surface waves with a DBT vortex in a controlled experiment.
The vortex is generated by pumping water at flow rate $Q$ into a rectangular ($2.65~\mathrm{m}\times1.38~\mathrm{m}$) tank and allowing it to drain through a hole of radius $d=2~\mathrm{cm}$ located in the centre.
Once the vortex is in equilibrium (determined by the constancy of $Q$ and $H_0$) monochromatic surface waves are generated using a series of electrically controlled pistons.
The change in water height $\Delta H=H-H_0$ is determined by illuminating the free surface with a laser sheet and tracking it's average position with a high-speed camera.
More details of the method can be found in Appendix C.

In Table~I of Appendix C, we summarise the parameters used in each experiment. 
We display the measured $\Delta H(t)$ profiles in Fig.~\ref{fig:results}.
In experiment 1 (panel A) we verify the linear decrease of $H$ at early times predicted by Eq.~\eqref{Hlin}.
In experiment 2 (panel B) we test the frequency dependence of Eq.~\eqref{Hdot5} by varying the frequency from 2 to 4 Hz, finding no significant variation of $\dot{H}_0$ within this range.
In experiment 3 (panel C) we demonstrate that at late times the behaviour of $H(t)$ deviates from a linear decline.

For the results of the first two experiments, we fit $\Delta H(t)$ with the linear decline predicted in Eq.~\eqref{Hlin} to provide a value for $\dot{H}_0$.
Since we do not have a prediction for $H(t)$ over the full evolution, we devise a phenomenological model which has the expected behaviour at early and late times, i.e. linear then exponential,
\begin{equation} \label{sech2}
H(t) = H_0 + \tfrac{1}{2}\dot{H}_0(t+t_c) - \dot{H}_0\tau\log\big(2\cosh\big[\tfrac{t-t_c}{2\tau}\big]\big).
\end{equation}
The extra parameters are the time $t_c$ at which the evolution switches from linear to exponential, and the exponential decay time $\tau$.
Strictly speaking, this model has linear asymptotics in the limit $t\to-\infty$.
However, the behaviour at $t=0$ is well approximated as linear provided $t_c>\tau$.
The parameters obtained from the fits are also contained in Table~I.

\begin{figure*} [t!] 
\centering
\includegraphics[width=\linewidth]{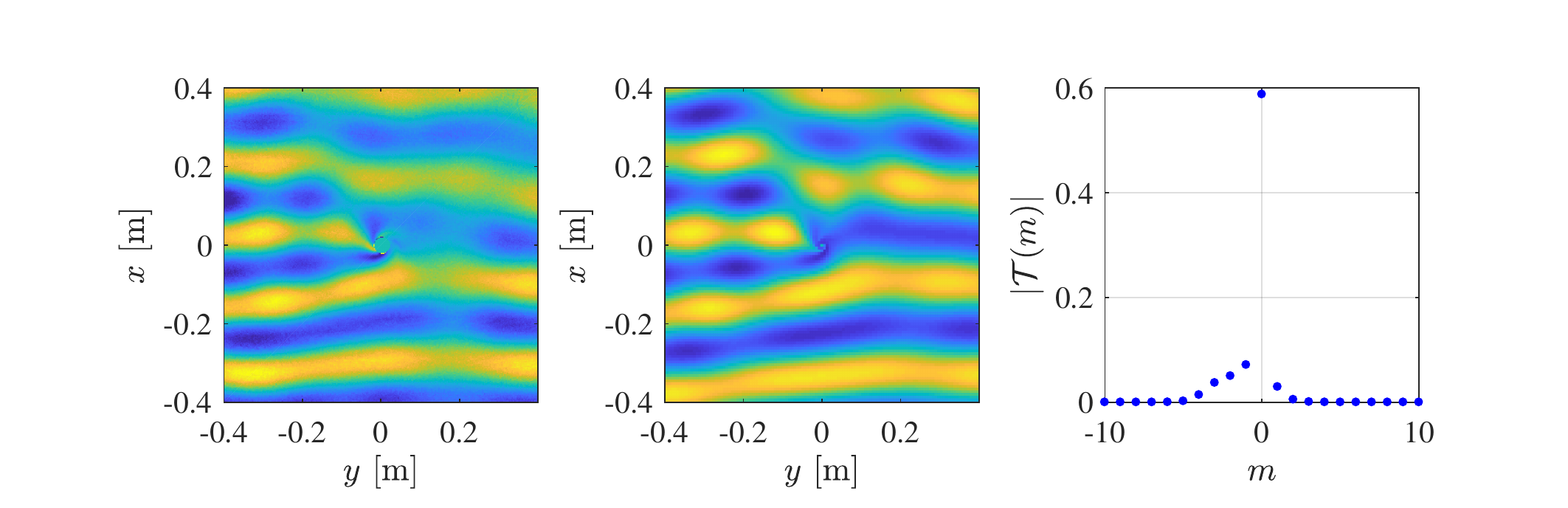}
\caption{Comparison between a scattered $f=2~\mathrm{Hz}$ wave produced under the conditions of experiment 2 (left) and the prediction from a simulation of the shallow water equations for an irrotational vortex with a flat free surface (centre).
A plane wave is generated at $x<0$ and propagates toward $x>0$. The vortex rotates counter-clockwise.
On the left, the curvature of the free surface near the drain is too large for our detection method to resolve, and hence we exclude data points for $r<d$.
The similarity between the two images is apparent, thereby supporting our simplified theoretical treatment.
We also display the transmission coefficients $\mathcal{T}_m$ (right) obtained from the simulation, which enter into the equation for $\dot{H}_0$ via the transmitted wave amplitude $|A_m^h|=\mathcal{T}_m\tfrac{ga}{\omega}(2\pi\omega/c)^{-1/2}$ (see Eq.~(B8) in Appendix B). Therefore, only the low lying $m$ modes (and mainly those with $m\leq 0$) contribute to $\Delta H$.} \label{fig:scatter}
\end{figure*}

\textbf{\emph{Discussion.}} 
Across all experiments, we find linear behaviour at early times, lending support to our prediction in Eq.~\eqref{Hlin}.
This behaviour persists whilst $\Delta H\ll H_0$ as expected, which can be seen by comparing $\Delta H$ in Fig.~\ref{fig:results} with the corresponding value of $H_0$ in Table~I.
The observed lack of frequency dependence in the second experiment is supported by computation of Eq.~\eqref{Hdot5} assuming a shallow water flow (see Appendix B), which predicts that the value $\dot{H}_0$ only varies by $\sim1\%$ over the range $f\in[2,4]~\mathrm{Hz}$.
Furthermore, by comparing the gradients across the different experiments we see that $\dot{H}_0$ depends on $H_0$, supporting our claim that the late time behaviour should be exponential.
This is further evidenced by the late time tail in experiment 3.

In Fig.~\ref{fig:scatter}, we compare a simulation of the shallow water wave equation (see Eq.~(B5) in Appendix B) for $H_0=2~\mathrm{cm}$ and $f=2~\mathrm{Hz}$ with the scattered wave from experiment 2 for the same frequency. 
A mathematical justification of our inviscid, shallow water treatment for this particular experiment can be found in Appendix C.
This treatment is corroborated by the clear similarities between experiment and simulation in Fig.~\ref{fig:scatter}.
The scattered wave in our set-up was measured using the air-water interface sensor described in \cite{torres2017rotational} and for the simulation, we used the flow parameters $C=0.013~\mathrm{m^2/s}$ and $D=0.001~\mathrm{m^2/s}$ with the velocity profile in Eq.~(B2) of Appendix B. 
These values were chosen to be similar to that found from the flow measurements of \cite{torres2017rotational,torres2018application}, which involved the same experimental apparatus, and the precise values were tuned to match the number of wavefronts on the left and right sides of the images \footnote{Changing the values of $C$ and $D$ by 10\% resulted in notable visual discrepancies between the two scattering patterns in Fig.~\ref{fig:scatter}. The predicted value of $\dot{H}_0$ changed by about 60\% when using these values.}.
Using the form of the perturbations on the horizon, which requires knowledge of the transmission coefficient (see Appendix B), Eq.~\eqref{Hdot4} predicts $\dot{H}_0=-0.011~\mathrm{mm/s}$.
Comparing with the measured value $\dot{H}_0=-9.8\pm0.2\times10^{-3}~\mathrm{mm/s}$ from in Table~I, we find good agreement, despite the violation of $H\neq H(r)$ and $\bm{\nabla}\times\mathbf{V}=0$ (also assumed by our prediction) in the vortex core where the horizon is located (see Appendix C for further discussion).

To relate the effect we have observed to the backreaction in GR, consider the following.
For slowly evolving background, one can adopt different timescales for the background and the fluctuations (i.e. a Born-Oppenheimer approximation \cite{born1927quantentheorie}). 
In this approximation, which is also employed to study the backreaction in GR \cite{balbinot2006hawking}, the fluctuations of a shallow water, irrotational fluid obey Eq.~\eqref{KleinGordon}, with the components of the evolving effective metric given by,
\begin{equation} \label{metric}
g_{\mu\nu} = \begin{pmatrix}
-(gH-V_{\perp,i}V_\perp^i) & -V_{\perp,i} \\
-V_{\perp,j} & \delta_{ij}
\end{pmatrix},
\end{equation}
where the indices $i$ and $j$ run over spatial dimensions, $\delta_{ij}$ is the Kronecker delta function and the $t$ dependence enters via $H(t)$ and $\mathbf{V}_\perp(t) = V_\perp^i(t)\mathbf{e}_i$.
In our analysis, we have estimated only $H(t)$ a short time after the beginning of wave incidence and the full dynamics of $H(t)$ and $\mathbf{V}_\perp(t)$ warrant further investigation.
Furthermore, since the energy and angular momentum density of the shallow water background flow are given by $\varepsilon = \tfrac{1}{2}H\mathbf{V}_\perp^2$ and $L = H\mathbf{r}\times\mathbf{V}_\perp$ respectively, a change in $H$ corresponds to changes in both $\varepsilon$ and $L$. Therefore, the height change mediates the exchange of energy and angular momentum between the waves and background.

Although the analogy persists at the linear level, the backreaction equation, describing how the effective metric evolves, is specific to the equations of motion of the system under consideration, which are the Euler equations for ideal fluids and the Einstein equations in GR (different backreaction equations have been studied, for example, in \cite{balbinot2005quantum,balbinot2006hawking,schutzhold2005quantum,schutzhold2007quantum}).
Despite this, it may still be possible to learn about generic features of slowly backreacting spacetime geometries using analogue systems. In addition, it is possible extract scattering amplitudes from changes in global parameters (here the water height) for controlled scattering processes, e.g.~scattering of waves with a single azimuthal wave number. 




\textbf{\emph{Conclusion.}} In this work, we have studied the backreaction of surface waves on a draining vortex flow.
Our results demonstrate that surface waves interacting with an initially stationary vortex will trigger the evolution of the background out of equilibrium. 
Due to the flow being externally driven, it was previously unclear whether the background had the freedom to adjust to the presence of waves in analogue gravity simulators. Our findings show that the backreaction is indeed observable, and that the system does in fact have freedom to re-distribute energy and angular momentum between the incident waves and the background flow.
In the shallow water regime, we have argued that this evolution is encapsulated by a dynamical effective metric. Although this metric does not evolve according to the Einstein equation, further similarities between slowly evolving gravitational and analogue spacetimes have yet to be investigated.

This realisation is important for a number of reasons. Firstly, one must ensure that any wave effects (e.g. stimulated Hawking radiation, superradiance and quasinormal ringing) are measured on a timescale much shorter than the time it takes for the background to change, so that the assumption of a stationary background is not violated. This may restrict the frequency range one can probe in an analogue gravity experiment, as non-linear effects will influence the low frequency behaviour which takes places over longer timescales. 
Secondly, the effect we have described is a global (as opposed to local) phenomenon. Thus, one can use the asymptotic value of the water height to obtain insight into scattering processes, similar to the role that the black hole mass plays in GR.

Based on previous experience, a similar behaviour is expected to occur in suitable quantum systems, and thus our findings suggest that analogue gravity experiments can be used to cross-validate backreaction models in a relativistic setting. This is an area of research where lack of experimental input is stalling theoretical development. \\


\acknowledgements
\textbf{\emph{Acknowledgements}}. SW acknowledges financial support provided under the Paper Enhancement Grant at the University of Nottingham, the Royal Society University Research Fellow (UF120112), the Nottingham Advanced Research Fellow (A2RHS2), the Royal Society Enhancement Grant (RGF/EA/180286) and the EPSRC Project Grant (EP/P00637X/1). SW acknowledges partial support from STFC consolidated grant No. ST/P000703/.  We also want to thank T.~Torres, S.~Erne, Z.~Fifer, A.~Geelmuyden and T.~Sotiriou for many useful discussions, as well as J.~Louko, W.~G.~Unruh and R.~M.~Wald for their insightful feedback.

\bibliographystyle{apsrev4-1}
\bibliography{backreaction_arXiv_2020_07_07.bbl}

\appendix
\numberwithin{equation}{section}
\section{}

\noindent Starting from first principles, we derive a general formula for the rate of change of total mass contained in a system due to a non-zero mass flux over it's boundary.
Consider a volume $V$ bounded by a surface $\mathcal{S}$.
In the absence of sources and sinks, a change in the density inside $V$ is compensated by the flow of mass over $\mathcal{S}$,
\begin{equation} \label{massbalance}
\int_V\partial_t\rho~dV = -\iint_\mathcal{S}\mathbf{J}\cdot d\mathbf{A} = -\int_V\bm{\nabla}\cdot(\rho\mathbf{V})dV,
\end{equation}
where in the second equality we have used the definition of the mass flux $\mathbf{J}=\rho\mathbf{V}$, as well as the divergence theorem to convert the surface integral into one over the volume.
The water tank we consider is bounded from below and open from above, $z\in[0,\infty)$, with water occupying the region $z\in[0,H]$ where $z=H$ is the free surface (or height) of the water.
Thus the density may be represented by $\rho=\rho_f\Theta(H-z)$, where $\rho_f=\mathrm{const}$ is the fluid's density and $\Theta$ is the Heaviside step function.
Inserting into Eq.~\eqref{massbalance} yields,
\begin{equation}
\int_\mathcal{A}\int^H_0\rho_f\bm{\nabla}\cdot\mathbf{V}dz~dA = 0.
\end{equation}
Here we have exploited the kinematic boundary condition at the free surface $V_z(z=H)=(\partial_t+\mathbf{V}_{\perp}\cdot\bm{\nabla}_{\perp})H$, where $\perp$ denotes the components perpendicular to the vertical coordinate $z$.
The area $\mathcal{A}$ of the system in the $\perp$ plane is assumed constant.
Splitting the divergence into $\perp$ and $z$ components, performing the $z$ integral and using the kinematic boundary condition, we arrive at,
\begin{equation} \label{Hdot1}
\begin{split}
\dot{M} = \rho_f\int_\mathcal{A}\partial_tH~dA = -\int_\mathcal{A}\Big[ \mathbf{J}_{\perp}\cdot\bm{\nabla}_{\perp}H - J_z(z=0) \\ + \int^H_0 \bm{\nabla}_{\perp}\cdot \mathbf{J}_{\perp} dz \Big] dA,
\end{split}
\end{equation}
where we have used the fact that the total mass of the system is given by $M=\rho_f\int_\mathcal{A}HdA$.
If the system is in equilibrium, all of the terms on the right hand side cancel leading to $\dot{M}=0$.

We can gain further intuition by considering the shallow water regime, where the expression for $\dot{M}$ simplifies significantly.
The shallow water approximation consists in assuming $L\ll H$, where $L$ is a length scale characterising variations of the fluid in the $\perp$ plane. 
In this approximation, the fluid is treated as a single two dimensional layer, in which the $z$ dependence in $\mathbf{V}_\perp$ is neglected.
Eq.~\eqref{Hdot1} then reduces to,
\begin{equation} \label{Hdot2}
\dot{M} = -\int_\mathcal{A}\bm{\nabla}_\perp\cdot(H\mathbf{J}_\perp)dA = -\int_\gamma H\mathbf{J}_\perp\cdot d\mathbf{l},
\end{equation}
where $\gamma$ is the boundary of $\mathcal{A}$ and we have also assumed that $J_z(z=0)=0$ inside $\mathcal{A}$ (this is motivated by the set-up we consider in the main text).
Thus, the system is in equilibrium if total mass flux over $\gamma$ is zero.

Now, let $(h,\mathbf{v})$ be linear fluctuations about the background, which are switched on at the time $t=0$.
The presence of fluctuations contributes additional terms to the mass flux.
Making the replacements,
\begin{equation}
H \to H + \epsilon h, \qquad \mathbf{V}\to \mathbf{V} + \epsilon\mathbf{v},
\end{equation}
where $\epsilon$ is an order tracking parameter, Eq.~\eqref{Hdot2} can be written,
\begin{equation} \label{Hdot3}
\begin{split}
\int_\mathcal{A}\Big[\partial_tH + & \bm{\nabla}_{\perp}\cdot(H\mathbf{V}_\perp+ \epsilon^2h\mathbf{v}_{\perp})\Big] dA \\ 
& = \epsilon \int_\mathcal{A}\Big[\partial_th + \bm{\nabla}_{\perp}\cdot(H\mathbf{v}_\perp + h\mathbf{V}_{\perp}) \Big] dA,
\end{split}
\end{equation}
where we have gathered the terms at linear order in $\epsilon$ on the right hand side.
Assuming that for $t<0$ background is in equilibrium, the presence of fluctuations will trigger the evolution of the background through the $\mathcal{O}(\epsilon^2)$ term in Eq.~\eqref{Hdot3}.
To separate out the background and fluctuation dynamics, once can introduce two new time variables $T_1 = t$ and $T_2 = \epsilon^2t$, such that $h=h(T_1)$, $H=H(T_2)$ and similarly for the velocity field. 
Furthermore, since $\bm{\nabla}_{\perp}\cdot(H\mathbf{V}_\perp)=0$ just before $t=0$, the leading order contribution to this term will also be $\mathcal{O}(\epsilon^2)$.
Requiring that the governing equations be satisfied to each order in $\epsilon$, the left and right hand sides of Eq.~\eqref{Hdot3} must both vanish respectively.

To compute change in $H$ close to $t=0$, we can drop the term $\bm{\nabla}_{\perp}\cdot(H\mathbf{V}_\perp)$ since the background is still approximately in equilibrium.
Assuming that $H$ is independent of spatial coordinates, the height change at early times is dictated by,
\begin{equation}
\dot{H}_0 \simeq -\frac{1}{\mathcal{A}}\int_\gamma h\mathbf{v}_\perp\cdot d\mathbf{l},
\end{equation}
which is the result stated in Eq.~(3) of the main text.
Note that the quantity under the integral is related to the Stokes drift velocity $\mathbf{v}_S$ for small waves in shallow water by $H\mathbf{v}_S=h\mathbf{v}_\perp$ \cite{philips1977dynamics}.
Thus, we may interpret the mechanism behind the height change as the Stokes drift integrated over the boundary of the system.
At later times, the $\bm{\nabla}_{\perp}\cdot(H\mathbf{V}_\perp)$ term in Eq.~\eqref{Hdot3} (which is replaced by the first and third term on the right hand side of Eq.~\eqref{Hdot1} for generic water depth) must also be taken into account, meaning one has to solve the full backreaction equation.

The result in Eq.~(3) is not specific to shallow water and only relies on the assumption of small waves, since the perturbed $z$ integral in Eq.~\eqref{Hdot1} from $H$ to $H+\epsilon h$ reduces to a multiplication, resulting in the same $\mathcal{O}(\epsilon^2)$ term as in Eq.~\eqref{Hdot3}.

\section{}

\noindent In this appendix, we deal with a $t$ independent background to derive $\dot{H}_0$ in Eq.~(3). We can make considerable progress analytically by assuming a shallow fluid which is also irrotational, i.e. $\bm{\nabla}\times\mathbf{V}=0$. 
This second assumption means that the velocity field represents only a single degree of freedom $\Phi$ satisfying $\mathbf{V}=\bm{\nabla}\Phi$.
Similarly, the fluctuations satisfy $\mathbf{v}=\bm{\nabla}\phi$.
The equations for the stationary background are,
\begin{equation} \label{SWeqns}
\frac{1}{2}\mathbf{V}_\perp^2 + gH = \mathrm{const}, \qquad \bm{\nabla}_\perp\cdot(H\mathbf{V}_\perp) = 0.
\end{equation}
The first of these is Bernoulli's equation resulting from momentum conservation, and the second comes from Eq.~\eqref{Hdot2}.
An approximate solution to Eq.~\eqref{SWeqns} far from the vortex core where $H\simeq\mathrm{const}$ is,
\begin{equation} \label{DBT}
\mathbf{V}_\perp = -D/r~\mathrm{\mathbf{e}}_r + C/r~\mathrm{\mathbf{e}}_\theta,
\end{equation}
which is known in the literature as the draining bathtub vortex (DBT), and $C>0$ and $D>0$ are the circulation and drain constants.
The perturbed shallow water equations are then combined into a wave equation for the fluctuations,
\begin{equation} \label{wave_equation}
(\partial_t+\mathbf{V}_\perp\cdot\bm{\nabla}_\perp)^2\phi - c^2\nabla^2_\perp\phi = 0,
\end{equation}
where $c=\sqrt{gH}$ is the wave speed. 
The system exhibits an acoustic horizon $r_h$ and ergosphere $r_e$ (defined respectively as the boundaries of the regions satisfying $|\mathrm{\mathbf{e}}_r\cdot\mathbf{V}|>c$ and $|\mathbf{V}|>c$) which are given by,
\begin{equation}
r_h = D/c, \qquad r_e = \sqrt{C^2+D^2}/c.
\end{equation}
Since the horizon is a one-way membrane for the perturbations, $r_1=r_h$ is a natural choice of inner boundary for the system.
The solution for the fluctuations on the horizon then determines the form of $\dot{H}_0$.
Using the separation ansatz in Eq.~(5) for the field $\phi$, Eq.~\eqref{wave_equation} becomes,
\begin{equation} \label{ODE}
\begin{split}
& (c^2r^2-D^2)r^2\partial_r^2\phi_m \\ 
+ & \left[D^2+c^2r^2-2iD\left(\omega r^2-mC\right)\right]r\partial_r\phi_m \\
+ & \left[(\omega r^2-mC)^2-2imCD-m^2c^2r^2\right]\phi_m = 0.
\end{split}
\end{equation}
This has a regular singular point at $r=r_h$ and thus a solution may be sought in terms of a Frobenius expansion about $r=r_h$ \cite{churilov2018scattering}.
Imposing that the solution remain finite, the first two terms of the series are,
\begin{equation} \label{frob}
\begin{split}
\phi_m(r) = A_m^h\left[1 - \frac{\tilde{\omega}^2-\frac{2imCD}{r_h^4}-\frac{m^2c^2}{r_h^2}}{2c\left(\frac{c}{r_h}-i\tilde{\omega}\right)}(r-r_h) + ... \right],
\end{split}
\end{equation} 
where $A_m^h$ and $\tilde{\omega}=\omega-mC/r_h^2$ are the mode amplitude and intrinsic angular frequency on the horizon respectively.
To find $h_m$ and $\mathbf{v}_m$, we use the relations,
\begin{equation}
\begin{split}
h_m(r_h) = & \ i\tilde{\omega}g^{-1}\phi_m(r_h) + cg^{-1}\partial_r\phi_m|_{r=r_h}, \\
\mathbf{v}_m(r_h) = & \ \partial_r\phi_m|_{r=r_h}\mathrm{\mathbf{e}}_r +imr_h^{-1}\phi_m(r_h)\mathrm{\mathbf{e}}_\theta,
\end{split}
\end{equation}
which upon insertion into Eq.~(6) yields,
\begin{equation} \label{Hdot6}
\dot{H}_0 = -\frac{\pi}{4gc\mathcal{A}}\sum_m F|A_m^h|^2,
\end{equation}
where $F$ (which controls the sign of $\dot{H}_0$) is a function of the flow and wave parameters given by,
\begin{equation}
F = \frac{\tilde{\omega}^4 - \frac{c^8}{D^4}\left(4\omega m\frac{C}{c^2}+m^4\right)}{\tilde{\omega}^2+\frac{c^4}{D^2}}.
\end{equation}
Note that for $|\tilde{\omega}|\gg c^2/D$ we have $F\simeq\tilde{\omega}^2$, which is the result one would naively expect from the eikonal approximation.
Unlike the energy current, $\dot{H}_0$ does not change sign with $\tilde{\omega}$. 
This is a result of the mass flux only being one of several terms contained in the energy current when $\mathbf{V}\neq0$ (see Appendix D for discussion).

\section{} 

\begin{table*}[t]
\small
\begin{center}
\begin{tabular}{c|c|c|c|c|c|c|c|c}
Exp. & $Q~[\mathrm{\ell/min}]$ & $H_0$~[mm] & $f$~[Hz] & $a$~[mm] & $\Delta t$~[s] & $\dot{H}_0$~[mm/s] &  $\tau$~[s] & $t_c$~[s] \\
\hline
$1$ & $14.0\pm0.4$ & $20\pm1$ & $4$ & $1.6\pm0.1$ & $120$ & $-0.0118\pm0.0002$ & - & - \\
\hline
$2$ & $14.0\pm0.4$ & $19\pm1$ & $2,3,4$ & $2.3\pm0.5$ & $120$ & $-0.0098\pm0.0002$ & - & - \\
\hline
$3$ & $29.4\pm0.4$ & $65\pm1$ & $4$ & $2.1\pm 0.5$ & $600$ & $-0.0389\pm0.0005$ & $94\pm4$ & $383\pm3$ \\
\end{tabular}
\end{center}
\caption{Details of the different experiments performed to measure the height change $\Delta H$. $Q$ is the average flow rate over the course of the experiment; $H_0$ is the initial water height; $f$ and $a$ are the incident wave frequency and amplitude respectively; $\Delta t$ is the window of wave incidence; $\dot{H}_0$ is the gradient of the linear height change at early times; $\tau$ is the decay of the exponential decrease at late times and $t_c$ is approximately the time when the behaviour switches from linear to exponential. The models used to extract these parameters are stated in in Eqs.~(4) and~(7). The uncertainties on the parameters are computed using the residuals from the fitting.}
\label{tab:exps}
\end{table*}

\noindent Here we expand on the details of our experiments.
To observe the height change resulting from impinging waves, we illuminate the free surface from above using a Yb-doped laser, mean wavelength $457~\mathrm{nm}$, which is converted into a thin laser sheet (thickness $\sim2~\mathrm{mm}$). This appears on the free surface as a line spanning nearly the full length of the tank, thereby allowing us to see the free surface in our data. We filmed this line from the side with a high-speed Phantom Miro Lab camera at $24~\unit{fps}$. The line of sight of the camera was at an angle of $\bar{\theta}$ to the free surface such that $\cos\bar{\theta}=0.91\pm0.01$, which was necessary to avoid shadows of waves passing in between the laser-sheet and the camera, which obscured our vision of the free surface. This induces an uncertainty on the measured height change which we include in our error estimates of $\dot{H}_0$. We recorded the height before sending any waves to confirm that the background was steady, and then monitored the water height whilst exciting waves of frequency $f=\omega/2\pi$ and amplitude $a$ over a time window $\Delta t$. The free surface was identified by finding the pixel of maximum intensity in each image and interpolating using adjacent points to determine the maximum to a sub pixel accuracy. 

Once the free surface in each image is determined as a function of spatial coordinate, we extract the zero wavenumber contribution to the signal using a spatial Fourier transform. This gave a measurement of the average height across the observed region at each time step, which we correct for the camera angle $\bar{\theta}$, obtaining the variation of the background height over time. In all experiments, the initial height was determined to be sufficiently steady (less than $10\%$ of the total change over the experiment) and any slight variation was due to the difficulty in maintaining constant $Q$ over the experiment. We corrected for this (as well as the slightly different initial heights) by fitting the curve prior to the start of the waves with a straight line and subtracting this from the entire data set to give the height change $\Delta H$. This ensured that the height change measured was the result of the perturbations.

In all experiments, we observed a small amplitude oscillation about the linear behaviour at the frequency $f$ (confirmed by a peak in the Fourier transform of $\Delta H$) corresponding the oscillatory (linear) terms in Eq.~\eqref{Hdot3}.
Furthermore, the amplitude of this oscillation decreased with $f$ which is expected, since integrating $\dot{H}_0$ in time brings out a factor $1/f$. We also observe a sharp increase in height immediately before the height starts to decrease. This is because our wave generator panel is initially fully retracted and must move forward to its equilibrium position, thereby reducing the area of the system and increasing mean height. In accordance with this explanation, a sharper increase was observed for larger piston amplitudes.

To justify a shallow water treatment, we make the following argument.
The dispersion relation is obtained from the wave equation by making the replacements $i\partial_t\to\omega$ and $-i\bm{\nabla}_\perp\to\mathbf{k}$, where $\mathbf{k}$ is the wavevector in the $\perp$ plane.
For $kH<1$, where $k=||\mathbf{k}||$, the dispersion relation for gravity waves \cite{landau1987fluid} may be expanded in powers of $kH$,
\begin{equation} \label{disp}
\begin{split}
\omega-\mathbf{V}_\perp\cdot\mathbf{k} = \pm ck(1 - k^2H^2/6 + ...~),
\end{split}
\end{equation}
where the first term on the right hand side corresponds to the wave equation in Eq.~\eqref{wave_equation}, and the second term is the leading order dispersive correction for $kH\neq 0$.
To estimate the size of this term far from the vortex core where $||\mathbf{V}_\perp||\ll c$, take $k\simeq\omega/c$, which is valid up to corrections of $\mathcal{O}((kH)^2)$ and $\mathcal{O}(||\mathbf{V}_\perp||/c)$.
Using this estimate, only experiment 2 for $f=2,3~\mathrm{Hz}$ satisfies $kH<1$.
For these frequencies, the dispersive correction in Eq.~\eqref{disp} is $0.05$ and $0.11$ respectively and thus, in particular for $f=2~\mathrm{Hz}$, our shallow water treatment seems reasonable.

A similar argument may be invoked to justify the assumption of an inviscid fluid.
Indeed, corrections to the dispersion relation due to viscosity $\nu$ appear in powers of $\nu k/c$ \cite{visser1998acoustic,ganguly2017acoustic}, which can be seen purely on dimensional grounds.
For water at room temperature with $\nu=1~\mathrm{mm^2/s}$, $\nu k/c\sim\mathcal{O}(10^{-5}f)$ and can thus be safely neglected for all frequencies considered.

Finally, although the assumptions $H\neq H(r)$ and $\bm{\nabla}\times\mathbf{V}=0$ are satisfied far from the centre, they are violated in the vortex core.
The effect of free surface gradients was studied in \cite{unruh2013irrotational,coutant2014undulations,richartz2015rotating} and it was shown that whilst the KG equation is preserved, $H=H(r)$ alters the form of the effective metric.
Flows with vorticity, however, do not preserve the KG equation since the usual scalar mode $\phi$ becomes coupled to additional degrees of freedom \cite{visser2004vorticity,liberati2019vorticity}.
Since both effects become important in the vortex core where the drain is located, their inclusion into into the description of the fluctuations would be necessary to improve the prediction for $\dot{H}_0$.
However, since both are satisfied far from the vortex core, the KG equation still holds in this region and we can still interpret wave propagation there using the analogy to GR.

\section{}

\noindent Here we provide a discussion of the different notions of energy current and their relation to the mass flux.
We begin with an action for the perturbations $\mathcal{S}=\int\mathcal{L}d^2xdt$, where the Lagrangian density is given by,
\begin{equation} \label{Lagrangian}
\begin{split}
\mathcal{L} = & \ -\tilde{g}^{\mu\nu}\partial_\mu\phi\partial_\nu\phi^* \\
= & \ |D_t\phi|^2-c^2|\bm{\nabla}_\perp\phi|^2,
\end{split}
\end{equation}
where $\tilde{g}^{\mu\nu} = \sqrt{-g}g^{\mu\nu}$ is the inverse metric and the squared absolute value is obtained by taking the product of a quantity with its complex conjugate. We have also defined the material derivative $D_t = \partial_t+\mathbf{V}_\perp\cdot\bm{\nabla}_\perp$ for conciseness. 
The explicit form of $\tilde{g}^{\mu\nu}$ is,
\begin{equation}
\begin{split}
\tilde{g}^{\mu\nu} = \ & \begin{pmatrix}
-1 & -V_\perp^i \\
-V_\perp^j & c^2\delta^{ij}-V_\perp^iV_\perp^j
\end{pmatrix} \\
= \ & -\delta^\mu_t\delta^\nu_t +(c^2\delta^{ij}-V_\perp^iV_\perp^j)\delta^\mu_i\delta^\nu_j - 2V_\perp^i\delta^\mu_{(i}\delta^\nu_{t)}.
\end{split}
\end{equation}
The wave equation in Eq.~\eqref{wave_equation} is then obtained using the field theoretic version of the Euler-Lagrange equations,
\begin{equation} \label{ELeqns}
\partial_\mu\left(\frac{\partial\mathcal{L}}{\partial(\partial_\mu\phi)}\right) - \frac{\partial\mathcal{L}}{\partial\phi} = 0.
\end{equation}
Note that applying Eq.~\eqref{ELeqns} to the first line of Eq.~\eqref{Lagrangian} results in the KG equation of Eq.~(1), whereas the second line yields directly the shallow water wave equation in Eq.~\eqref{wave_equation}.

A conserved current $j[\phi]$ is a quantity with components $j^{\mu}[\phi]$ satisfying,
\begin{equation}
\partial_{\mu}j^{\mu}[\phi] = 0 \quad \Rightarrow \quad \partial_t\rho[\phi]+\bm{\nabla}_\perp\cdot\mathbf{j}[\phi] = 0,
\end{equation}
where the second form in terms of $(\rho,\mathbf{j})$ splits $j$ into the temporal and spatial parts which are the charge $\rho$ and current $\mathbf{j}$ respectively (not to be confused with the density and mass flux in the main text). Note, the two together are collectively called the 4-current. Square brackets are use to indicate that a quantity is a functional of $\phi$.

Conserved currents are derived as follows. Consider an infinitesimal transformation of the field which induces a shift in the Lagrangian,
\begin{equation}
\phi^a\to\phi^a+\delta\phi^a \quad \Rightarrow \quad \mathcal{L}\to\mathcal{L}+\delta\mathcal{L}.
\end{equation}
Noether's theorem \cite{srednicki2007quantum,schwartz2014quantum} states that $j$ is a conserved current if the Lagrangian changes by a total derivative $\delta\mathcal{L}=\partial_\mu F^\mu$. The components of the current are given by,
\begin{equation}\label{NoetherCurrent}
\begin{split}
j^\mu = \ & \frac{\partial\mathcal{L}}{\partial(\partial_\mu\phi^a)}\delta\phi^a - F^\mu \\
= \ & -\tilde{g}^{\mu\nu}(\partial_\nu\phi\delta\phi^*+\partial_\nu\phi^*\delta\phi) - F^\mu,
\end{split}
\end{equation}
where $a=1,2$ with $\phi_1=\phi$ and $\phi_2=\phi^*$. We now consider two different conserved quantities: the norm and the energy.

\textit{Norm conservation}. We consider first performing a phase rotation on $\phi$. Since the wave equation is linear in $\phi$, there is an internal symmetry $\phi\to\exp(i\alpha)\phi$, where $\alpha$ is a phase rotation, which leaves the equations of motion invariant. 
For infinitesimal $\alpha$, $\phi$ changes by $\delta\phi=i\alpha\phi$ and $\delta\mathcal{L}=0$. The conserved quantities associated with this transformation are the norm $\rho_n[\phi]$ and the norm current $\mathbf{j}_n[\phi]$ given by,
\begin{equation} \label{n_current}
\begin{split}
j^\mu_n[\phi] = & \ i\tilde{g}^{\mu\nu}(\phi^*\partial_\nu\phi-\phi\partial_\nu\phi^*) \\
\qquad \quad \mathrm{or} \\
\rho_n[\phi] = & \ i(\phi^*D_t\phi-\phi D_t\phi^*) \\
\mathbf{j}_n[\phi] = & \ \ i\mathbf{V}_\perp(\phi^*D_t\phi-\phi D_t\phi^*) \\
& \qquad - ic^2(\phi^*\bm{\nabla}_\perp\phi-\phi\bm{\nabla}_\perp\phi^*).
\end{split}
\end{equation}
Using the ansatz in Eq.~(5) for an axisymmetric, quasistationary system, we have $\partial_t\rho_n=0$. Applying the divergence theorem to \mbox{$\bm{\nabla}_\perp\cdot\mathbf{j}_n[\phi] = 0$} over the region $r=[r_h,\infty)$, we deduce \mbox{$\int^{2\pi}_0[\mathbf{r}\cdot\mathbf{j}_n]^\infty_{r_h}d\theta = 0$}. Using asymptotic solutions to Eq.~\eqref{ODE} (see e.g. \cite{berti2004qnm}), this integral yields,
\begin{equation} \label{eq.amplitudes}
\omega(|A_m^+|^2-|A_m^-|^2) = -\tilde{\omega}|A_m^h|^2
\end{equation}
where the left (right) of the equation is the norm current expressed at infinity (the horizon) and $+/-$ denote out/in-going respectively.

\textit{Energy conservation}. Time translation $t\to t-\delta t$ induces a change in the field $\delta\phi^a=\delta t\partial_t\phi^a$ and the Lagrangian $\delta\mathcal{L}=\delta t\partial_t\mathcal{L}$. This gives rise to conservation of the energy current, which has components,
\begin{equation} \label{e_current}
\begin{split}
j^\mu_e[\phi] = & \ -\tilde{g}^{\mu\nu}(\partial_t\phi^*\partial_\nu\phi+\partial_t\phi\partial_\nu\phi^*) \\
\qquad \quad \mathrm{or} \\
\rho_e[\phi] = & \ \partial_t\phi^*D_t\phi + \partial_t\phi D_t\phi^* - \mathcal{L} \\
= & \ |\partial_t\phi|^2 + c^2|\bm{\nabla}_\perp\phi|^2-|\mathbf{V}_\perp\cdot\bm{\nabla}_\perp\phi|^2 \\
\mathbf{j}_e[\phi] = & \ \mathbf{V}_\perp(\partial_t\phi^*D_t\phi + \partial_t\phi D_t\phi^*) \\
& \qquad - c^2(\partial_t\phi^*\bm{\nabla}_\perp\phi + \partial_t\phi\bm{\nabla}_\perp\phi^*).
\end{split}
\end{equation}
Defining $\partial_t\phi=\dot{\phi}$ and $p=\partial\mathcal{L}/\partial\dot{\phi} = D_t\phi^{*}$ (and similarly for the complex conjugate), one can see that $\rho_e=\mathcal{H}$ is the Hamiltonian density. 

To make contact with fluid dynamics, we define \mbox{$(E,\mathbf{I})=(\rho_e,\mathbf{j}_e)/2g$} and also $\phi\in\mathbb{R}$. Using $D_t\phi=-gh$ and $\bm{\nabla}_\perp\phi=\mathbf{v}_\perp$, we obtain the usual equation for wave energy conservation in shallow water,
\begin{equation}
\partial_tE + \bm{\nabla}_\perp\cdot\mathbf{I}=0,
\end{equation}
where,
\begin{subequations}
\begin{align}
E = \ & \frac{1}{2}gh^2 + \frac{1}{2}H\mathbf{v}_\perp^2 + h\mathbf{V}_\perp\cdot\mathbf{v}_\perp, \label{eq.energydensity} \\
\mathbf{I} = \ & (h\mathbf{V}_\perp+H\mathbf{v}_\perp)(gh+\mathbf{V}_\perp\cdot\mathbf{v}_\perp). \label{eq.energycurrent}
\end{align}
\end{subequations}
In fluid dynamics, these equations are obtained by contracting the linear shallow water equations with $(gh,H\mathbf{v}_\perp^T)$ where superscript $T$ indicates the transpose. Following the same procedure outlined above, we evaluate the expression for $E$ to show $\partial_tE=0$, and applying the divergence theorem we obtain \mbox{$\int^{2\pi}_0[\mathbf{r}\cdot\mathbf{I}]^\infty_{r_h}d\theta = 0$}. Evaluating this expression in both limits results again in Eq.~\eqref{eq.amplitudes} multiplied by a factor of $\omega c/g$, thus making contact between the norm current and the energy current.

Restricting our analysis to $\omega>0$, notice that if $|A_m^+|^2<|A_m^-|^2$ the energy current is negative (points towards $r=r_h$). However if $|A_m^+|^2>|A_m^-|^2$ (which occurs for $\tilde{\omega}<0$) the energy current becomes positive and points outward, allowing for the extraction of energy from the system to infinity (this is the phenomenon of superradiant scattering). This is understood as the mode carrying a negative energy across the horizon, $\rho_e(r_h)<0$, thereby lowering the energy of the system.

As a brief aside, Eq.~\eqref{e_current} applied to harmonic field modes (of the form in Eq.~(5)) becomes,
\begin{align}
j^\mu_e[\phi]=i \omega \tilde{g}^{\mu\nu}\left(\phi^*\partial_\nu\phi-\phi\partial_\nu \phi^*\right).
\end{align} 
One can then observe that this energy $4$-current is $\omega$ multiplied by the norm $4$-current in \eqref{n_current}. Moreover, Eq.~\eqref{NoetherCurrent} applied to angular coordinate translations leads to an angular momentum $4$-current $j^\mu_l$ satisfying the well-known property $j^\mu_e/j^\mu_l=\omega/m$.

Now to make contact with the mass flux, we notice that when $\mathbf{V}=0$ we have $\mathbf{I}=gHh\mathbf{v}_\perp$. Once integrated, this is precisely the term appearing in Eq.~(3) multiplied by a factor $c^2$. 
Therefore, in the absence of a flow, there is a direct correspondence between the energy current in Eq.~\eqref{eq.energycurrent} and the shallow mass flux $\mathbf{j}_\mathrm{SW}=h\mathbf{v}_\perp$ given by $\mathbf{I}=\mathbf{j}_\mathrm{SW}c^2$ (reminiscent of the celebrated result $E=mc^2$). 
However, when $\mathbf{V}\neq0$, we see clearly from Eq.~\eqref{eq.energycurrent} that the energy current and mass flux no longer coincide. This explains why Eq.~(6) does not have the symmetry properties of $\tilde{\omega}$.

For the reader who is more familiar with the language of electromagnetism, we draw some comparisons here. The combination $\tfrac{1}{2}\mathrm{Re}[h_m^*\mathbf{v}_m]$ appearing in Eq.~(6) is reminiscent of the complex Poynting vector, $\mathbf{S}$, whose real part describes the flux of power in electromagnetic waves, in terms of the phasor decomposition of the electric and magnetic fields ($\mathbf{E}$ and $\mathbf{H}=\mathbf{B}/\mu_0$ respectively), \cite{meystre2007classical},
\begin{equation}
\mathbf{S}=\frac{1}{2}\mathbf{E}\times\mathbf{H}^*.
\end{equation}
The correspondence between $\mathbf{I}$ and $\mathbf{S}$ becomes exact upon defining,
\begin{align*}
\frac{h}{H} = \frac{\mathbf{B}}{B_0}\cdot\mathrm{\mathbf{e}}_z, \qquad \frac{\mathbf{v}_\perp}{c} = \frac{\mathbf{E}}{E_0}\times\mathrm{\mathbf{e}}_z,
\end{align*}
where $E_0=cB_0$ are reference electric/magnetic fields respectively and the correspondence requires $H\to\varepsilon_0$ and $gH\to1/\mu_0\varepsilon$. Indeed, one can show using this correspondence that Maxwell's equations in the vacuum are equivalent to the shallow water system of equations in an irrotational, quiescent ($\mathbf{V}=0$) fluid. 
Hence, in standing water, the mass flux is analogous to the electromagnetic power flux.

\end{document}